\documentclass[sigconf]{acmart}
\AtBeginDocument{%
  \providecommand\BibTeX{{%
    \normalfont B\kern-0.5em{\scshape i\kern-0.25em b}\kern-0.8em\TeX}}}

\usepackage{geometry}
\usepackage{longtable}
\usepackage{multirow}
\usepackage{booktabs}
\usepackage{makecell}
\usepackage{verbatim}

\setcopyright{acmcopyright}
\copyrightyear{2025}
\acmYear{2025}
\acmDOI{XXXXXXX.XXXXXXX}

\acmConference[CHI '25]{Proceedings of the CHI Conference on Human Factors in Computing Systems}{April 26--May 1, 2025}{Yokohama, Japan}
\acmISBN{978-1-4503-XXXX-X/18/06}

\begin{document}

\title[Mentigo: Intelligent Agent for Mentoring Students in the CPS]{Mentigo: An Intelligent Agent for Mentoring Students in the Creative Problem Solving Process}


\author{Siyu Zha}
\orcid{1234-5678-9012}
\authornote{The authors contributed equally to this research.}
\affiliation{%
  \institution{Tsinghua University}
  \city{Beijing}
  \country{China}}
  \email{zhasiyu22@mails.tsinghua.edu.cn}

\author{Yujia Liu}
\authornotemark[1]
\affiliation{%
  \institution{Tsinghua Univeristy}
  \city{Beijing}
  \country{China}}
  \email{l-yj22@mails.tsinghua.edu.cn}

\author{Chengbo Zheng}
\affiliation{%
  \institution{Hong Kong University of Science and Technology}
  \city{Hongkong}
  \country{China}}
  \email{cb.zheng@connect.ust.hk}

\author{Jiaqi XU}
\affiliation{%
 \institution{University of California Los Angeles}
 \city{California}
 \country{USA}}
 \email{beatrice15@g.ucla.edu}

\author{Fuze Yu}
\affiliation{%
 \institution{Beijing University of Technology}
 \city{Beijing}
 \country{China}}
 \email{yufuze@emails.bjut.edu.cn}

 \author{Jiangtao Gong}
 \authornote{Corresponding Author}
\orcid{0000-0002-4310-1894}
\affiliation{%
  \institution{Tsinghua Univeristy}
  \city{Beijing}
  \country{China}}
   \email{gongjiangtao2@gmail.com}

 \author{Yingqing XU}
\affiliation{%
  \institution{Tsinghua Univeristy}
  \city{Beijing}
  \country{China}}
   \email{yqxu@tsinghua.edu.cn}

\renewcommand{\shortauthors}{Zha, et al.}

\begin{abstract}
With the increasing integration of large lauguage models (LLMs) in education, there is growing interest in using AI agents to support student learning in creative tasks. This study presents an interactive Mentor Agent system named Mentigo, which is designed to assist middle school students in the creative problem solving (CPS) process. We created a comprehensive dataset of real classroom interactions between students and mentors, which include the structured CPS task management, diverse guidance techniques, personalized feedback mechanisms. Based on this dataset, we create agentic workflow for the Mentigo system. The system's effectiveness was evaluated through a comparative experiment with 12 students and reviewed by five expert teachers. The Mentigo system demonstrated significant improvements in student engagement and creative outcomes. The findings provide design implications for leveraging LLMs to support CPS and offer insights into the application of AI mentor agents in educational contexts.

\end{abstract}



\maketitle

\section{Introduction}
Generative Artificial Intelligence (GAI) applications, such as ChatGPT, have attracted significant attention from educators and researchers worldwide~\cite{hwang2023exploring}. Although concerns about its misuse persist, there is no doubt that GAI has become one of the most transformative technologies in recent years~\cite{salas2022artificial}. Scholars have highlighted the potential of using GAI in education to improve student computational, creative and critical thinking skills~\cite{bahroun2023transforming}. However, research on how to effectively integrate GAI into different educational contexts and engage middle school students in real-world problem-solving is still in its early stages~\cite{leaton2020artificial,woithe2023understanding,rudolph2023chatgpt}.

Creative Problem Solving (CPS) is a well-established instructional method that encourages students to develop creativity by working collaboratively on complex real-world problems~\cite{samson2015fostering}. Empirical studies have shown that CPS can significantly enhance learning engagement, stimulate creativity, and promote active learning among students~\cite{samson2015fostering,puryear2020defining}. However, implementing CPS in educational settings also presents several challenges. Teachers must provide continuous and tailored guidance to ensure educational coherence and quality, especially when managing diverse interdisciplinary project topics~\cite{eberle2021cps}. These challenges are more pronounced in middle school settings~\cite{park2012development}. 
These challenges drive the demand for increasing the number of mentors to facilitate the CPS process of students, which, however, is a challenge for most schools due to the intensive teaching resources needed~\cite{mcfadzean2002developing}.

GAI presents an opportunity to address the challenges mentioned above in CPS. To this end, in this paper, we explore GAI acting as a mentor, providing real-time feedback and scaffolding CPS tasks~\cite{gizzi2022creative,neumann2020intelligent,ashoori2007mentor}. They can potentially support the CPS process by guiding students through each stage, thereby reducing the burden on human teachers and allowing for more scalable implementations. Recent discussions in the human-computer interaction (HCI) and educational technology communities have focused on exploring how GAI applications can be integrated into learning designs to facilitate CPS~\cite{lo2023impact,abd2023large,perkins2023academic}. Our key novelty is that, while previous work has primarily positioned GAI as a tool for task assistance, we explore GAI as a mentor that guides, supports, and fosters deeper learning and skill development aligned with students' real needs in educational settings, particularly for middle school CPS projects. Furthermore, there is limited research examining the application of GAI in middle school settings from an evidence-based perspective, highlighting the need for more targeted investigations.

To address these gaps, this study introduces Mentigo, an interactive AI-driven system designed to mentor middle school students throughout the CPS process. Mentigo combines structured guidance, adaptive feedback, and empathetic interactions to provide tailored support for students. The system operates through a six-stage CPS process, where it sets clear task objectives and progression standards at each stage. Mentigo dynamically adapts its mentoring strategies by monitoring students' cognitive and emotional states, utilizing a comprehensive database that matches these states with contextually relevant guidance strategies. This database-driven approach enables Mentigo to deliver personalized, flexible educational experiences that are responsive to the unique needs of each learner, fostering deeper engagement and critical thinking during the CPS tasks. We conducted a user study to evaluate Mentigo's effectiveness compared to a baseline system. The study showed that Mentigo significantly enhanced student engagement and promoted deeper cognitive processes, such as analysis and creativity, by providing structured guidance and adaptive feedback. Participants noted that the system's ability to offer personalized, reflective guidance improved their understanding and application of CPS skills.  

In summary, the contributions of this study are as follows:
\begin{itemize} 
\item Development of Mentigo, an interactive system that supports a human-AI collaborative CPS process, empowering students at each stage of CPS in middle school; 
\item Collection of real-world datasets from middle school students engaged in CPS, providing valuable insights into student-mentor interactions; 
\item Implementation of a comparative experiment and an expert review study to evaluate the effectiveness of Mentigo in fostering CPS skills; 
\item Design implications derived from the design process and user studies, offering guidelines for supporting student CPS with generative AI in educational contexts. \end{itemize}

Overall, this study provides a valuable exploration of the application of AI mentor agents and offers insights for future research on their use in educational settings.

\section{Related Work}

\subsection{Students‘ Creative Problem-Solving}
Creative Problem-Solving (CPS) is a powerful teaching method that can support a pedagogical shift in the classroom and foster both student engagement and motivation to learn ~\cite{samson2015fostering}. CPS enables students to tackle everyday problems using both creative and critical thinking skills, which are essential for developing flexible and adaptive learners ~\cite{eberle2021cps}. 
While many current studies focus on creative expression through open-ended activities such as drawing or storytelling, creativity in education is increasingly associated with problem-solving and finding innovative solutions~\cite{puryear2020defining}. 
Within the framework of CPS, creativity is defined as a process by which individuals or groups can formulate problems, generate and evaluate diverse and novel options, and effectively implement new solutions or courses of action~\cite{treffinger1995creative}. Ali et al. further emphasized creativity within CPS as involving the generation of fluent, novel, and valuable ideas to solve problems~\cite{ali2022escape}.

Many studies have demonstrated that creativity can be enhanced through cognitive and computational approaches in design-based problem-solving scenarios~\cite{egan2016human}. CPS has also been explored extensively within the HCI community. Most of research focuses on providing digital aids to assist users in accomplishing creative tasks~\cite{10.1145/3290605.3300619}. 
The CPS process typically involves several key stages, including problem understanding, idea generation, and action planning~\cite{treffinger1995creative}. Gizzi et al. proposed a CPS framework comprising four main components: 1) problem formulation, 2) knowledge representation, 3) method of knowledge manipulation, and 4) method of evaluation~\cite{gizzi2022creative}. Treffinger et al. introduced a contemporary CPS framework (CPS Version 6.1TM) that includes stages such as understanding the challenge, generating ideas, preparing for action, and planning approaches. The "understanding the challenge" stage involves constructing opportunities, exploring data, and framing problems~\cite{treffinger2003creative}. In this study, we adopt a six-stage CPS process: "Problem Discovery," "Information Collection," "Problem Definition," "Solution Ideation," "Solution Evaluation," and "Solution Implementation."

Despite its potential, implementing CPS projects in classrooms presents several challenges. CPS requires students to develop engagement, innovation, and critical reasoning abilities as they collaboratively investigate and solve problems, acquiring necessary knowledge and skills along the way~\cite{samson2015fostering}. Moreover, teachers face the challenge of guiding students in group work, planning, problem-solving, idea testing, and peer presentations, often within constrained timeframes~\cite{mcfadzean2002developing}. Various roles support CPS tasks, including tutors, mentors, and teaching assistants (TAs), each with distinct approaches. Tutors focus on helping students master specific academic content or skills and facilitate an effective learning environment within small groups~\cite{jung2005becoming}. TAs assist with planning, preparing, and supporting teaching activities, such as grading and classroom management~\cite{kitpo2020learning}. In contrast, this study adopts the concept of mentoring, which provides comprehensive support through learner-centered relationships often involving counseling, friendship, and socialization~\cite{bush1999developing,hansford2006principalship,lee2003student}. Researchers advocate for mentoring in teaching to empower students to engage more deeply in CPS tasks, allowing both mentor and mentee to develop CPS skills together~\cite{lavender2006creative, yu2017mentoring}. Mentoring has also been widely recognized for enhancing both teacher development and student learning~\cite{hudson2013mentoring}.

Providing adequate human mentorship in classroom settings can be challenging due to resource constraints. AI offers a promising alternative for filling this gap~\cite{gizzi2022creative}. Intelligent Tutoring Systems (ITSs) have demonstrated consistent learning gains over decades. Recent studies are exploring the role of mentor agents in learning environments~\cite{neumann2020intelligent,ashoori2007mentor}. However, there is still a lack of research on the application of generative AI specifically in the context of CPS. Our study seeks to address this gap by designing a mentor agent to guide students through the CPS process. Through user studies, we aim to explore how an AI agent can aid students in solving real-world problems and developing their CPS skills.

\subsection{Generative AI for Student's Learning}
The adoption of generative AI in education is progressing cautiously, with methodologies for their effective use still emerging. Recent studies highlight the need for more structured strategies and guidelines for integrating generative AI into educational settings ~\cite{salas2022artificial,leaton2020artificial,rudolph2023chatgpt}. This reflects a broader trend where the deployment of generative AI in education remains a work in progress. 

Researchers' views on incorporating generative AI into educational settings vary widely. Some advocate for their potential to transform learning environments, citing tools like ChatGPT as instrumental in enriching education, supporting educators, and redefining objectives and practices ~\cite{lo2023impact,abd2023large,perkins2023academic,alqahtani2023emergent}. For instance, researchers highlighted generative AI could empower students' learning by providing personalized information, resources, and step-by-step solutions based on students’ unique learning needs and interests~\cite{kasneci2023chatgpt}. On the other hand, there are growing concerns regarding the potential misuse of generative AI in educational settings. Susnjak et al. revealed that essays generated by ChatGPT display advanced critical thinking and textual discourse, sparking debates about the risks of cheating in online assessments ~\cite{susnjak2022chatgpt}. These findings all underscore the implications and the need for careful oversight when using generative AI. Despite differing opinions, generative AI is still recognized as a vital support in educational instruction, indicating a significant role for these technologies in future learning\cite{vazquez2023chatgpt,nguyen2023ethical}. 

Besides, many researchers proposed that generative AI  can effectively assist in curriculum exercises to empower students' CPS ~\cite{tan2023towards,aydin2023chatgpt}. Recent studies are highlighting LLMs, particularly GPT-4, showing that they can be powerful tools to support CPS~\cite{hadi2023survey,chakrabarty2023creativity}. Noy et al. found that ChatGPT can enhance the quality of creative solutions~\cite{noy2023experimental}. Hwang, A. H. C., \& Won, A. S. investigated how human subjects collaborate with a computer-mediated chatbot in creative idea-generation tasks~\cite{hwang2021ideabot}. Douglas et al.'s research points out that LLMs can do even better than traditional brainstorming methods and average human performance in specific tasks when given the right prompts and paradigms ~\cite{summers2023brainstorm,yao2023tree}. Therefore, researchers recommended designing AI-involved learning tasks that engage students in real-world problem-solving to meet learning objectives effectively~\cite{woithe2023understanding}.  

In conclusion, generative AI has demonstrated the potential to improve students’ CPS. However, their actual use, especially involving generative AI in the CPS process, remains relatively rare. Our study aims to design an agent for mentoring students in the CPS process, with a focus on middle school students. Our goal is to offer insights into how generative AI can be effectively incorporated into educational settings, empowering students in each stage of CPS in middle school.


\section{Formative Interview \& Design Goals}
To better understand the challenges faced by educators and students in CPS tasks, we conducted a formative study consisting of in-depth interviews with experienced teachers and mentors. The goal of this study was to identify key areas where support is needed, thereby informing the design of an AI-based mentoring agent to assist in CPS process. The interviews provided valuable insights into the need for timely feedback, diversified guidance strategies, personalized support, and sustained motivation mechanisms in educational settings.

\subsection{Participants and Procedure}
Five experienced educators participated in this formative study, including three secondary school teachers who are in charge of CPS projects at their schools and two university mentors with backgrounds in education and extensive experience in mentoring middle school students. The three teachers have rich experience in designing and implementing CPS in their respective schools, while the two mentors have participated in various CPS projects for middle school students and have gained insights into student guidance.

The interviews were conducted via Tencent Meeting. The process began by collecting basic background information from participants, including their teaching or mentoring experience, years of practice, and current roles. During the interviews, the teachers were asked about their experiences in different phases of CPS implementation, including the specific strategies they employed, challenges encountered, and their use of AI tools to assist in teaching. We also explored their perspectives on how an AI agent could potentially support these projects in the future. Mentors were further asked about their experiences in guiding students, focusing on the types of questions commonly posed by students, the feedback they provide, and the challenges they observe in student performance. Additionally, they offered suggestions for improving guided strategies to better support students in CPS tasks.

\subsection{Findings}
\label{sec3-findings}
The findings below summarize the key areas where support and improvements are needed to enhance the teaching and learning experience in the CPS process.

\textbf{Need for Structured Guidance:}
Teachers emphasized the importance of providing structured guidance to help students navigate complex CPS tasks with a clear sense of direction and understanding of each stage's specific objectives. They noted that students need explicit instructions that clearly outline the tasks and expectations for each phase to ensure focused progress. Teachers also pointed out that it is challenging for them to constantly check in with each student to monitor their understanding and task alignment, making a structured and well-defined guidance system crucial for effective learning.

\textbf{Importance for Diverse Strategies:}
The study emphasized the need for a variety of guidance strategies to accommodate different learning styles and needs. They suggested incorporating multiple approaches, such as task planning, creative stimulation, and in-depth analysis, to effectively engage students and enhance their CPS skills. The interviewed mentors highlighted the effectiveness of scaffolding techniques and open-ended questions to encourage independent thinking and learning. Since CPS involves both divergent and convergent thinking stages, it is crucial for teachers to provide different types of guidance tailored to each phase to help students progressively acquire CPS-related knowledge and skills. Additionally, mentors pointed out the importance of specific feedback mechanisms and continuous evaluation to better support students' growth and development throughout the process.

\textbf{Value of Personalized Feedback:}
The study highlighted the importance of providing personalized feedback based on real-time student performance, as feedback tailored to a student's specific progress and needs is more effective in promoting learning and sustaining engagement. Teachers noted that students often require timely and detailed feedback throughout each stage of their CPS tasks, relying on precise, process-oriented guidance to navigate complex challenges. However, due to limitations in time and resources, it is difficult for teachers to deliver one-on-one personalized feedback to every student, particularly on professional and procedural aspects. This challenge is even more pronounced for novice teachers, who may struggle to provide effective guidance. In this context, educators suggested introducing a supportive role, such as a mentor, to offer timely and personalized feedback at various stages of project implementation.

\textbf{Encouraging Deep Cognitive Engagement:}
Teachers emphasized the importance of encouraging students to engage in deeper cognitive processes and critical thinking. They suggested that the educational agent should prompt students to explore problems thoroughly, consider various perspectives, and reflect on their reasoning to develop comprehensive solutions. Teachers face significant difficulties in prompting students to engage in deeper thinking, especially in helping them evaluate and justify the value and significance of their proposed solutions. Teachers widely believe that a mentor, providing one-on-one feedback, guidance, and direction, would greatly improve the teaching effectiveness in CPS tasks.

\textbf{Need for Emotional Support:}
Teachers emphasized the necessity of providing emotional support to students throughout CPS tasks. They highlighted that empathetic interactions, encouragement, and an understanding of students' emotional states are vital for maintaining motivation and fostering a positive learning environment. Educators pointed out that students often face frustration and anxiety during complex problem-solving activities, which can hinder their learning progress. Therefore, incorporating emotional support through a mentor who can recognize emotional cues and respond with empathy could help sustain students' engagement and confidence, ultimately contributing to a more supportive and effective educational experience.

These findings outline the various needs of students in CPS tasks and emphasize the importance of mentors in providing effective support to address these needs in the design of agents.

\subsection{Design Goals}
Based on the findings from our formative study (as detailed in \autoref{sec3-findings}), we establish five design goals (DGs) essential for developing an educational agent that mentors students in CPS tasks.

\textbf{DG1: Ensure Structured Task Management for Clarity and Compliance} In CPS, middle school students need structured task management and phased guidance to understand the specific requirements and expected outcomes at each stage~\cite{maier2018efficient}. Our findings emphasize the need for clear and explicit task guidance to help students stay focused and aligned with learning objectives. Therefore, the educational agent must incorporate rigorous task management features that clearly define and reinforce the objectives and requirements for each CPS phase~\cite{chang2020agent}. This structured approach will help students adhere to their learning goals and improve overall CPS performance by reducing ambiguity and enhancing task clarity.

\textbf{DG2: Diversify Guided Strategies for Enhanced CPS Skills} Effective CPS requires students to engage in deep, creative, and critical thinking, which is best supported through diverse guided strategies~\cite{treffinger2007creative,lim2020development}. Our findings suggest that employing a variety of pedagogical approaches, such as scaffolding techniques, open-ended questioning, and problem-based learning, can help students develop robust knowledge frameworks and tackle complex problems. Therefore, the educational agent should be designed to utilize a wide range of guided strategies that are adaptable and tailored to different problem-solving contexts, ensuring that guidance is dynamic and responsive to students' evolving needs.

\textbf{DG3: Provide Personalize Feedback through Dynamic Student State} 
Students require timely and personalized feedback throughout their CPS tasks, where they heavily depend on specific, process-oriented guidance~\cite{tetzlaff2021developing,basadur1990identifying}. However, teachers often struggle to provide one-on-one personalized feedback due to time and resource constraints. To address this need, the educational agent should develop mechanisms that continuously assess and interpret the student's states, enabling the delivery of contextually appropriate and individualized feedback. This approach would enhance the student learning experience and support their continuous development.

\textbf{DG4: Encourage Deep Cognitive Engagement and Conceptual Mastery} Maintaining student motivation during challenging creative tasks is essential for fostering deep cognitive engagement and achieving conceptual mastery~\cite{samson2015fostering,schroth2023spirituality}. Research indicates that students benefit from consistent encouragement to engage in critical thinking and reflective practices. Therefore, the educational agent should be designed to actively promote deep cognitive engagement by encouraging continuous reflection and critical analysis, guiding students to internalize and effectively apply CPS methodologies. This would help sustain interest, foster deeper understanding, and enhance long-term retention of concepts.

\textbf{DG5: Integrate Empathetic Interactions to Foster Emotional Engagement} Maintaining student motivation and emotional well-being is crucial, particularly during prolonged creative tasks where frustration and disengagement can occur~\cite{mccurdy2020problem}. Studies show that students benefit from empathetic support that addresses both their emotional and academic needs. To create a supportive learning environment, the educational agent should incorporate emotionally intelligent responses that simulate human-like empathy, helping to nurture both the student's emotional resilience and academic success. These design goals, grounded in research findings, provide a comprehensive framework for developing an educational agent that effectively supports students in CPS tasks through dynamic, adaptive, and empathetic learning interventions.

\section{The Mentigo Agent System}
\begin{figure*}[!htp]
    \centering
    \includegraphics[width=1.0\linewidth]{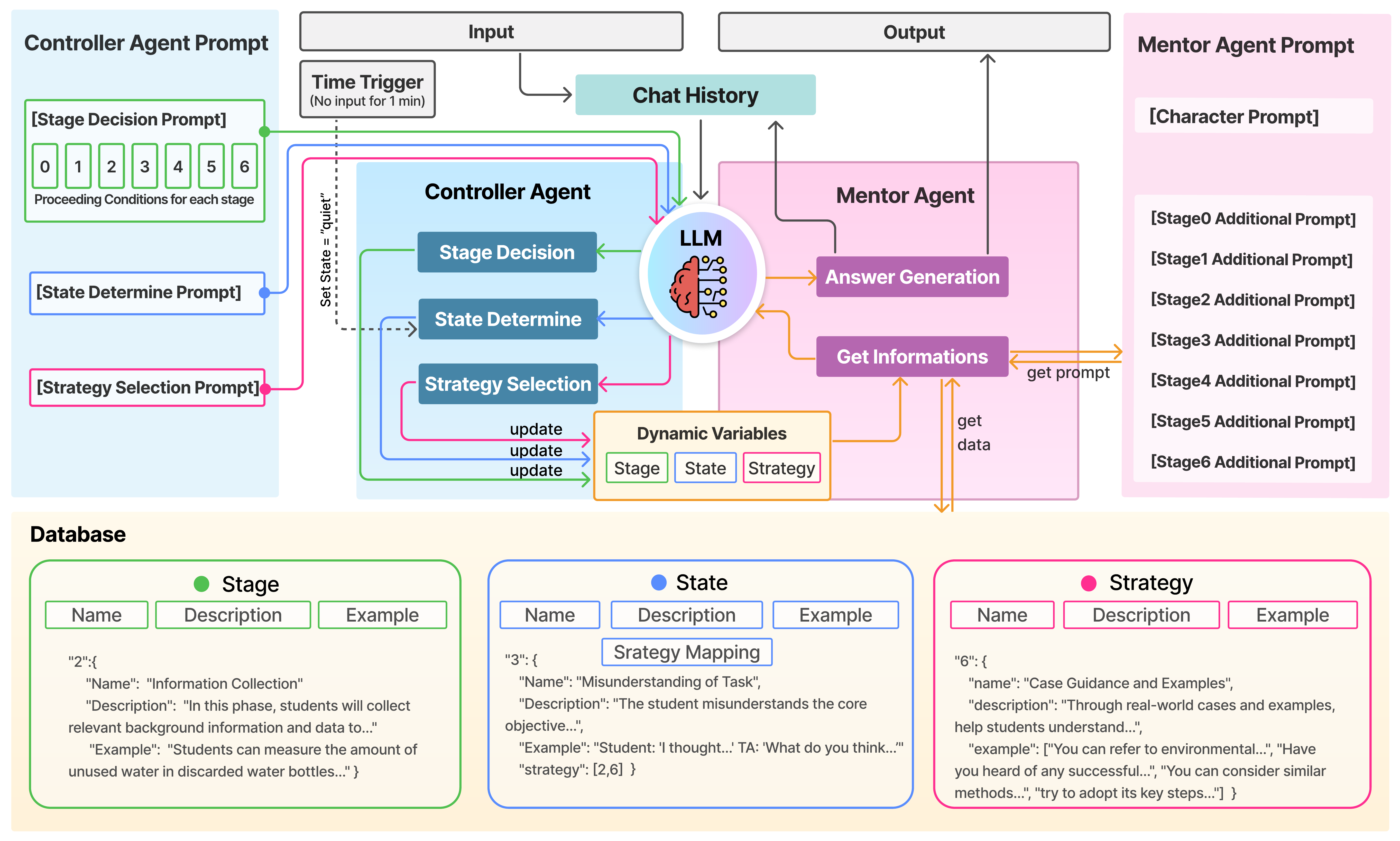}
    \caption{Framework of Mentigo Agent.The Controller Agent manages the flow of student interactions through three functions: Stage Decision, State Determine, and Strategy Selection. The Mentor Agent provides tailored feedback and guidance prompts based on students' progress. The Database stores information on different Stages, States, and Strategies to support adaptive decision-making, with all interactions facilitated by an LLM.
}
    \label{fig:framework}
\end{figure*}


In this section, we present the design of \textbf{Mentigo}, an LLM-based agent framework developed to act as a personalized mentor for students working on CPS tasks. The system framework (see Fig.~\ref{fig:framework}) comprises three main components: a database, a controller agent, and a mentor agent. To achieve the first three design goals (DG1-DG3), we collected and structured data from human mentors, which forms the foundation of the database in three parts: stages, states, and strategies. This mentor data allows the system to understand different CPS stages, recognize student states, and select appropriate guidance strategies, ensuring the agent delivers contextually relevant and effective mentoring aligned with authentic educational practices.

To accomplish the fourth and fifth design goals (DG4-DG5), we designed an agent workflow that integrates both a controller agent and a mentor agent. The controller agent is responsible for logical decision-making, including matching CPS stages, student states, and suitable guidance strategies, while the mentor agent focuses on generating empathetic interactions and controlling emotional expressions to enhance the mentoring experience. By combining the LLM’s advanced language capabilities with a curated database of human mentoring examples, \textbf{Mentigo} provides structured guidance, personalized feedback, diverse strategies, and empathetic support, enabling middle school students to navigate CPS tasks effectively. Finally, we outline the system implementation, including the dataset construction derived from authentic student-mentor interactions and the technical details that integrate these elements into a cohesive educational agent framework.

\subsection{Mentigo Dataset Construction}
This study focuses on extracting student states and mentor guidance strategies from authentic CPS scenarios to inform the design of our agent system. We conducted two CPS workshops involving 57 middle school students, divided into six groups per workshop, each guided by one of 12 mentors with backgrounds in education, technological development, or engineering. These workshops replicated classroom dynamics to capture interactions across the six stages of CPS. A total of 80 hours of classroom sessions were recorded, transcribed, and coded, resulting in 1,615 dialogue entries and 747 instances of guidance strategies and student responses. Using established frameworks, we systematically categorized the student states (e.g., confusion, confidence) and mentor strategies (e.g., probing, encouragement). This coding process produced a comprehensive dataset that captures the subtleties of real classroom interactions. This dataset serves as a foundational element for developing our agent system, enabling it to understand various student states and apply appropriate mentoring strategies. Based on this dataset, we achieve the first three design goals (DG1-DG3) as follows:

\subsubsection{[DG1]\textbf{Ensure Structured Task Management for Clarity and Compliance}}

\begin{table*}[h!]
\centering
\small
\caption{Stages of Creative Problem Solving (CPS)}
\begin{tabular}{|l|p{12cm}|}
\hline
\textbf{CPS Stages} & \textbf{Definition} \\ \hline
Problem Discovery & In this stage, students identify potential problems or challenges related to the project theme. Through observing the campus environment, discussions, or guidance, students can find practical issues worth attention. The goal of this stage is to help students discover opportunities for improvement related to the project theme around them. \\ \hline
Information Collection & In this stage, students gather relevant background information and data to better understand the problems they have discovered. They may conduct surveys, collect data, or talk to relevant individuals to obtain information related to the project theme. \\ \hline
Problem Definition & Based on the data and information collected, students define the core problem. This stage helps students clarify what needs to be solved and ensures they focus on the truly significant challenges. \\ \hline
Solution Ideation & Students generate multiple potential solutions through brainstorming and creative thinking. They are encouraged to propose a variety of innovative ideas, regardless of their initial practicality. \\ \hline
Solution Evaluation & In this stage, students evaluate the solutions they have generated and select the most effective and feasible one. They will consider factors such as feasibility, cost, potential impact, and how to gain more support and participation for the solution. \\ \hline
Solution Implementation & In this stage, students complete a final implementation plan report, which should include an introduction to the background of the chosen solution, the conception of the solution, a specific implementation plan, and potential challenges that may arise during the implementation process. \\ \hline
\end{tabular}
\label{tab:CPS_stages}
\end{table*}

To mentor students through CPS tasks clearly and effectively, the Mentigo system employs a structured task management approach using dynamic learning phases. The system divides the learning process into six distinct phases: "Problem Discovery," "Information Collection," "Problem Definition," "Solution Ideation," "Solution Evaluation," and "Solution Implementation." Each stage has clearly defined tasks, descriptions, and expected outcomes managed through a predefined "Stage".

This structure is part of the underlying dataset that supports the controller agent’s logical decision-making. Table 1 provides a comprehensive overview of each phase and its definitions, which serves as the foundation for the system to manage task progression and ensure that students are guided step-by-step. Grounded in educational theories that emphasize phased learning for complex tasks, this structured approach helps minimize cognitive overload and ensures students gain a thorough understanding of each phase's objectives. Maintaining focus on learning goals enhances task compliance and overall student performance.

\subsubsection{[DG2]\textbf{Diversify Guided Strategies for Enhanced CPS Skills}}

The system incorporates a range of guided strategies to provide adaptive learning support and enhance CPS skills. This approach is informed by pedagogical frameworks such as scaffolding and cognitive tools, which are known to promote deeper engagement and critical thinking in learners. Using a predefined "Stage," the system dynamically selects appropriate guided strategies based on students' cognitive and emotional states, ensuring that the guidance provided is contextually relevant and effective. 

To systematically develop a robust set of guided strategies for CPS tasks, we employed a structured coding process based on authentic classroom interactions. We began by identifying a broad range of potential strategies from recorded dialogues between mentors and students during various CPS stages. Three researchers independently coded these dialogues to extract specific mentoring strategies, resulting in 20 distinct guided strategies that reflect the diverse techniques used by mentors to support student learning. Each strategy was then clearly defined to capture its unique characteristics and application.

These 20 strategies were subsequently organized into five overarching categories: "Task Planning and Execution," "Creative Thinking Stimulation," "In-depth Thinking and Systematic Analysis," "Information Integration and Resource Support," and "Emotional Support and Motivation." This categorization provides a comprehensive framework for understanding how different strategies can be applied effectively in CPS contexts. Each strategy was also aligned with specific student states identified earlier, enabling targeted interventions. The strategy vocabulary was further refined through feedback from experienced mentors and educational experts to ensure it is practically relevant. The final set of 20 strategies, grouped into five major categories, is detailed in Table 2 and serves as the foundation for developing our system, allowing it to simulate effective mentoring by dynamically applying strategies based on student states and learning needs.

\begin{table*}[h!]
\small
\caption{Taxonomy of teachers' strategies for guiding students in CPS}
\begin{tabular}{|c|l|l|}
\hline
\textbf{ID} & \multicolumn{1}{c|}{\textbf{Guidance Category}}                        & \multicolumn{1}{c|}{\textbf{Specific Strategy}} \\ \hline
1           & \multirow{2}{*}{\textbf{Task Planning and Execution Support}}          & Task Follow-up and Allocation                   \\ \cline{1-1} \cline{3-3} 
2           &                                                                        & Task Restatement                                \\ \hline
3           & \multirow{5}{*}{\textbf{Creative Thinking Stimulation}}                & Brainstorming Guidance                          \\ \cline{1-1} \cline{3-3} 
4           &                                                                        & Random Association and Creative Stimulation     \\ \cline{1-1} \cline{3-3} 
5           &                                                                        & Creative Filtering and Focusing                 \\ \cline{1-1} \cline{3-3} 
6           &                                                                        & Case Guidance and Examples                      \\ \cline{1-1} \cline{3-3} 
7           &                                                                        & Challenge Guidance                              \\ \hline
8           & \multirow{5}{*}{\textbf{Deep Thinking and Systematic Analysis}}        & Probing Guidance                                \\ \cline{1-1} \cline{3-3} 
9           &                                                                        & System Thinking Training                        \\ \cline{1-1} \cline{3-3} 
10          &                                                                        & Critical Thinking Practice                      \\ \cline{1-1} \cline{3-3} 
11          &                                                                        & Deep Thinking and Reflection                    \\ \cline{1-1} \cline{3-3} 
12          &                                                                        & Rebuttal and Debate                             \\ \hline
13          & \multirow{5}{*}{\textbf{Information Integration and Resource Support}} & Resource Guidance                               \\ \cline{1-1} \cline{3-3} 
14          &                                                                        & Content Review                                  \\ \cline{1-1} \cline{3-3} 
15          &                                                                        & Information Collection and Classification       \\ \cline{1-1} \cline{3-3} 
16          &                                                                        & Information Integration Guidance                \\ \cline{1-1} \cline{3-3} 
17          &                                                                        & Scenario Simulation and Feedback                \\ \hline
18          & \multirow{3}{*}{\textbf{Emotional Support and Motivation Stimulation}} & Encouragement and Motivation                    \\ \cline{1-1} \cline{3-3} 
19          &                                                                        & Emotional Counseling                            \\ \cline{1-1} \cline{3-3} 
20          &                                                                        & Immediate Feedback                              \\ \hline
\end{tabular}
\end{table*}

\subsubsection{[DG3]\textbf{Provide Personalize Feedback through Dynamic Student State}}

Mentigo continuously monitors students' progress to deliver personalized feedback by interpreting real-time inputs related to engagement, performance, and emotional states. Drawing from our real-world dataset analysis, we identified common student states in CPS tasks and matched them with appropriate guidance strategies. For instance, if a student is "stuck," the system provides "case-based guidance," whereas for a student "lacking confidence," it offers encouraging feedback.

\begin{table*}[h!]
\small
\caption{Students' states in CPS and potential teacher guidance strategies}
\begin{tabular}{|c|c|l|l|}
\hline
\textbf{ID} & \textbf{Student Status}                                                                                                              & \multicolumn{1}{c|}{\textbf{Specific Status}}                                       & \multicolumn{1}{c|}{\textbf{Corresponding Guidance Strategies}}                                                                                                                                  \\ \hline
1           & \multirow{4}{*}{\textbf{\begin{tabular}[c]{@{}c@{}}Task/ Goal \\ Definition Unclear\end{tabular}}}                                   & Silent                                                                              & Task Assignment, Restate Task                                                                                                                                                                    \\ \cline{1-1} \cline{3-4} 
2           &                                                                                                                                      & Goal Unclear                                                                        & Task Assignment, Restate Task, Probing Guidance                                                                                                                                                  \\ \cline{1-1} \cline{3-4} 
3           &                                                                                                                                      & Task Misunderstanding                                                               & Restate Task, Case Guidance and Examples                                                                                                                                                         \\ \cline{1-1} \cline{3-4} 
4           &                                                                                                                                      & Goal Prioritization Confusion                                                       & Task Assignment, Restate Task                                                                                                                                                                    \\ \hline
5           & \multirow{2}{*}{\textbf{\begin{tabular}[c]{@{}c@{}}Time Management\\ and Task\\ Breakdown Difficulties\end{tabular}}}                & \begin{tabular}[c]{@{}l@{}}Poor Time\\ Management\end{tabular}                      & Task Assignment, Resource Guidance                                                                                                                                                               \\ \cline{1-1} \cline{3-4} 
6           &                                                                                                                                      & \begin{tabular}[c]{@{}l@{}}No Ideas/Task\\ Breakdown Difficulties\end{tabular}      & \begin{tabular}[c]{@{}l@{}}Brainstorming Guidance, Encouragement\\ and Motivation, Immediate Feedback\end{tabular}                                                                               \\ \hline
7           & \multirow{3}{*}{\textbf{\begin{tabular}[c]{@{}c@{}}Lack of Creative\\ and\\ Divergent Thinking\end{tabular}}}                        & Limited Thinking                                                                    & \begin{tabular}[c]{@{}l@{}}Creative Flow Selection and Focus,\\ Random Association and Creative\\ Stimulation, Scenario Simulation\\ and Scene Feedback, Case Guidance and Examples\end{tabular} \\ \cline{1-1} \cline{3-4} 
8           &                                                                                                                                      & Insufficient Ideas                                                                  & \begin{tabular}[c]{@{}l@{}}Random Association and Creative Stimulation,\\ Encouragement and Motivation,\\ Deep Thinking and Reflection\end{tabular}                                              \\ \cline{1-1} \cline{3-4} 
9           &                                                                                                                                      & Overly Divergent Thinking                                                           & \begin{tabular}[c]{@{}l@{}}Creative Flow Selection and Focus,\\ Refutation and Debate, Systematic Thinking Training\end{tabular}                                                                 \\ \hline
10          & \multirow{3}{*}{\textbf{\begin{tabular}[c]{@{}c@{}}Insufficient\\ In-depth\\ Thinking and\\ Systematic\\ Analysis\end{tabular}}}     & Shallow Thinking                                                                    & \begin{tabular}[c]{@{}l@{}}Challenge Guidance, Systematic\\ Thinking Training, Probing Guidance\end{tabular}                                                                                     \\ \cline{1-1} \cline{3-4} 
11          &                                                                                                                                      & Lack of Critical Thinking                                                           & \begin{tabular}[c]{@{}l@{}}Critical Thinking Exercises, Refutation and Debate,\\ Scenario Simulation and Scene Feedback\end{tabular}                                                             \\ \cline{1-1} \cline{3-4} 
12          &                                                                                                                                      & Lack of Systematic Thinking                                                         & \begin{tabular}[c]{@{}l@{}}Systematic Thinking Training, Deep Thinking\\ and Reflection, Critical Thinking Exercises\end{tabular}                                                                \\ \hline
13          & \multirow{3}{*}{\textbf{\begin{tabular}[c]{@{}c@{}}Insufficient\\ Information\\ Collection\\ and Integration\\ Skills\end{tabular}}} & \begin{tabular}[c]{@{}l@{}}Inadequate Information\\ Collection\end{tabular}         & \begin{tabular}[c]{@{}l@{}}Information Gathering and Classification,\\ Resource Guidance, Content Review\end{tabular}                                                                            \\ \cline{1-1} \cline{3-4} 
14          &                                                                                                                                      & \begin{tabular}[c]{@{}l@{}}Inadequate Information\\ Analysis Skills\end{tabular}    & \begin{tabular}[c]{@{}l@{}}Information Integration Guidance,\\ Refutation and Debate, Case Guidance and Examples\end{tabular}                                                                    \\ \cline{1-1} \cline{3-4} 
15          &                                                                                                                                      & \begin{tabular}[c]{@{}l@{}}Inadequate Information\\ Integration Skills\end{tabular} & \begin{tabular}[c]{@{}l@{}}Information Integration Guidance, Systematic\\ Thinking Training, Probing Guidance\end{tabular}                                                                       \\ \hline
16          & \multirow{2}{*}{\textbf{\begin{tabular}[c]{@{}c@{}}Evaluation and\\ Decision-Making\\ Difficulties\end{tabular}}}                    & Lack of Evaluation Criteria                                                         & \begin{tabular}[c]{@{}l@{}}Case Guidance and Examples, Critical Thinking\\ Exercises, Deep Thinking and Reflection\end{tabular}                                                                  \\ \cline{1-1} \cline{3-4} 
17          &                                                                                                                                      & Decision Uncertainty                                                                & \begin{tabular}[c]{@{}l@{}}Resource Guidance, Task Assignment,\\ Systematic Thinking Training\end{tabular}                                                                                       \\ \hline
18          & \multirow{2}{*}{\textbf{\begin{tabular}[c]{@{}c@{}}Lack of \\ Motivation\\ and Confidence\end{tabular}}}                             & Insufficient Intrinsic Motivation                                                   & Encouragement and Motivation, Immediate Feedback                                                                                                                                                 \\ \cline{1-1} \cline{3-4} 
19          &                                                                                                                                      & Lack of Confidence                                                                  & \begin{tabular}[c]{@{}l@{}}Encouragement and Motivation, Emotional\\ Support, Case Guidance and Examples\end{tabular}                                                                            \\ \hline
20          & \multirow{4}{*}{\textbf{\begin{tabular}[c]{@{}c@{}}Fatigue/ Burnout/ \\ Negative Emotions\end{tabular}}}                             & Learning Fatigue                                                                    & Encouragement and Motivation, Emotional Support                                                                                                                                                  \\ \cline{1-1} \cline{3-4} 
21          &                                                                                                                                      & Low Mood                                                                            & Encouragement and Motivation, Emotional Support                                                                                                                                                  \\ \cline{1-1} \cline{3-4} 
22          &                                                                                                                                      & Conflicts Arising                                                                   & \begin{tabular}[c]{@{}l@{}}Intergroup Encouragement,\\ Role Guidance, Immediate Feedback\end{tabular}                                                                                            \\ \cline{1-1} \cline{3-4} 
23          &                                                                                                                                      & \begin{tabular}[c]{@{}l@{}}Normal Progress/\\ Active Questioning\end{tabular}       & Encouragement and Motivation, Immediate Feedback                                                                                                                                                 \\ \hline
\end{tabular}
\end{table*}

We developed a comprehensive set of student state codes to analyze learning states in CPS scenarios. The process began with an initial screening of classroom recordings to identify relevant student states associated with CPS activities. Three researchers independently coded the dialogues, resulting in 23 specific student states that represent a range of behaviors and emotional responses during CPS tasks. These states were refined into eight broader categories, such as "Time Management Challenges" and "Emotional Support Needs," providing a clearer framework for understanding student engagement.

Each student state is linked to corresponding guidance strategies to address specific needs, making the dataset both descriptive and actionable. For example, the state "Confused Goals" is associated with strategies like task clarification. The vocabulary was further refined with input from mentors and educational experts to ensure it captures the full spectrum of student behaviors. The final dataset, comprising 23 student states categorized into eight major categories with definitions and associated strategies (see Table 3 for details), serves as a foundation for the agent system to identify student states and apply the appropriate guidance strategies during CPS activities.

\subsection{Mentor Agent Workflow}

\subsubsection{[DG4]\textbf{Encourage Deep Cognitive Engagement and Conceptual Mastery}}

To better manage the CPS process and counteract the potential instability of large language models, we developed a controller agent that uses structured outputs to strictly manage three dynamic parameters: stage, state, and strategy. The controller agent operates solely in a decision-making capacity without direct interaction with users, ensuring consistency and clarity in guiding students through each phase of the CPS process.

\begin{figure*}[!htp]
    \centering
    \includegraphics[width=1.0\linewidth]{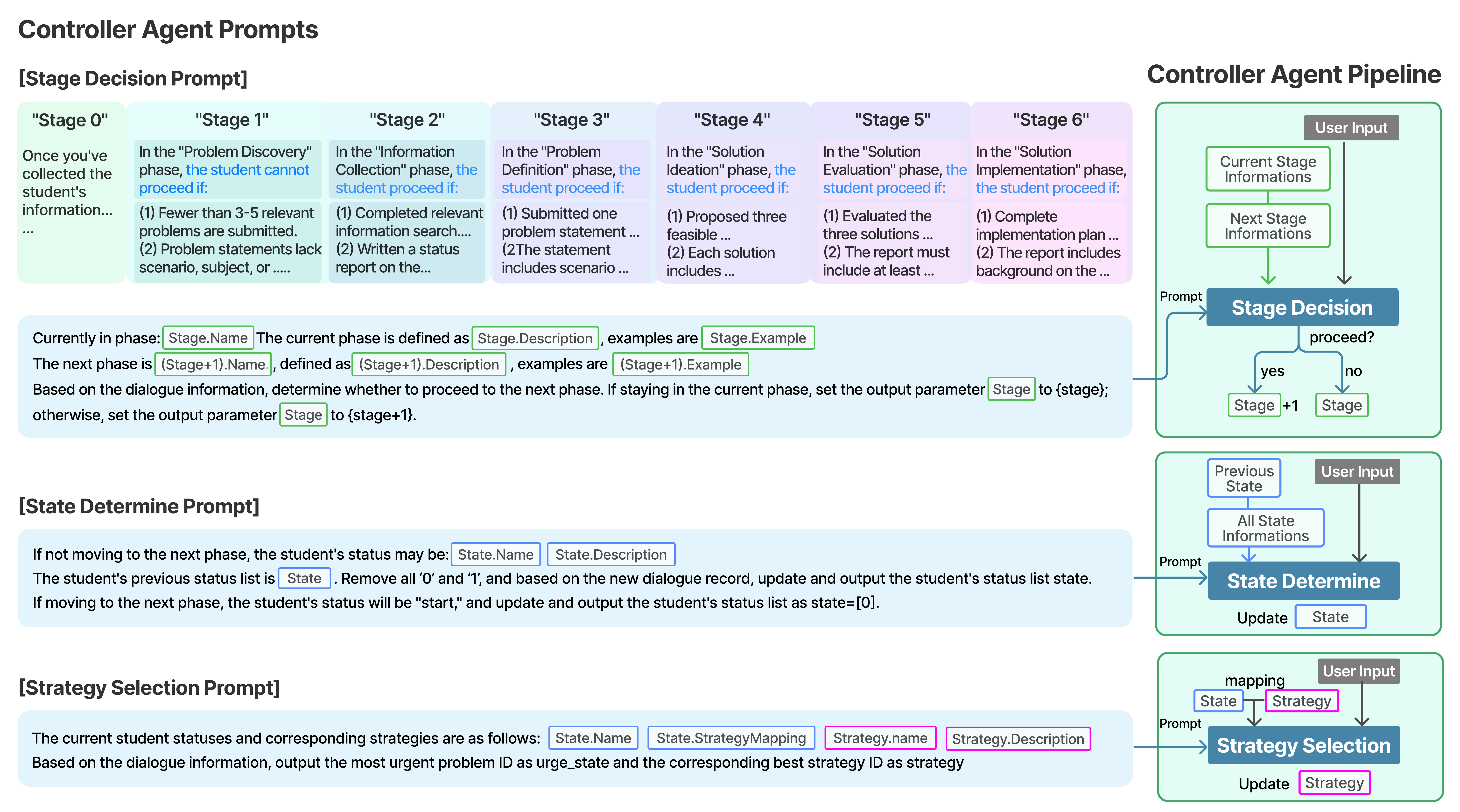}
    \caption{Prompts for Controller Agent}
    \label{fig:Controller}
\end{figure*}

The system begins by prompting the large model for a "stage decision" after each dialogue round. Each stage has strict criteria that must be met (see Fig.~\ref{fig:Controller}, Stage Decision Prompt), and only when all requirements are fulfilled can the process advance to the next stage (stage +1). Otherwise, the process remains in the current stage, and the "stage" parameter is updated accordingly. This method ensures that students fully comprehend each phase's requirements before moving forward.

Simultaneously, the system determines the "student state" by analyzing real-time inputs. If the stage advances, the state is updated to “stage start.” If not, the system provides the model with possible student states, their definitions, and the chat history to assess the student's current state(s), excluding “stage start” and “quiet” states. The "quiet" state is a special condition triggered by inactivity (e.g., no user input for one minute) to encourage proactive guidance. The updated "state" parameter list is then used to identify which strategies are most suitable for the student's current situation.

Once the stage and state parameters are updated based on the current dialogue, the system maps each state to corresponding guidance strategies. The model autonomously determines which student state requires immediate attention and selects the most appropriate strategy for that state. This process results in a dynamically chosen "strategy" that promotes deep cognitive engagement and supports students in mastering CPS concepts. This adaptive approach, grounded in real-time assessment and structured prompts, ensures that the agent remains responsive to individual student needs and provides targeted, context-aware guidance.

\subsubsection{[DG5]\textbf{Integrate Empathetic Interactions to Foster Emotional Engagement}}

\begin{figure*}[!htp]
    \centering
    \includegraphics[width=1.0\linewidth]{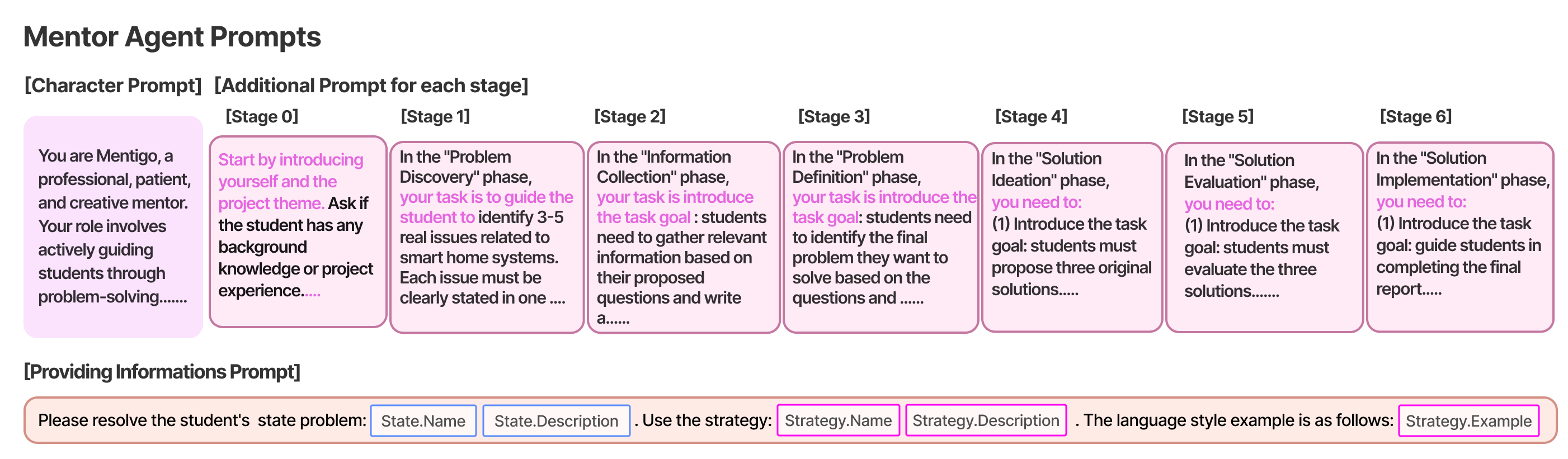}
    \caption{Prompts for Mentor Agent}
    \label{fig:mentor}
\end{figure*}

To enhance emotional engagement and create a more supportive learning environment, we designed a mentor agent specifically to guide students empathetically through their CPS tasks, as shown in Fig.~\ref{fig:mentor}. The mentor agent is structured around two key components: the "character prompt" and the "additional prompt for each step." The "character prompt" defines the mentor agent's overall persona, tone, and communication style, providing a consistent framework for interaction that is input into the system at every dialogue round. This persona is designed to be approachable, supportive, and motivational, creating a positive emotional connection with the student.

The "additional prompt for each step" serves as a supplementary guide tailored to each phase of the CPS process, offering more specific instructions and examples that align with the current stage's objectives. These prompts help the mentor agent provide relevant and context-aware guidance that encourages students to stay engaged, reflect deeply, and think critically about their tasks. By adjusting the prompts according to the requirements of each phase, the system ensures that the mentoring is both effective and aligned with pedagogical goals.

Moreover, we aim for Mentigo to dynamically adjust its guidance strategies based on the student's evolving states, as determined by the Controller Agent. Using the parameters provided by the Controller Agent—such as the current stage, student state, and chosen strategy—the system accesses the database to retrieve appropriate strategies and example dialogues. These examples are crafted to make the mentor agent’s responses more precise, concise, and natural, mirroring how a human mentor would communicate. All relevant information is then input into the mentor agent, which synthesizes this data to generate the final output, ensuring that each interaction is empathetic and tailored to the student's immediate needs. This design allows the mentor agent to not only guide students effectively but also to build an emotional rapport that fosters deeper engagement and motivation throughout the learning process.

\subsection{System Implementation}

\subsubsection{Dataset Construction}

From the 80 hours of classroom recordings involving 57 middle school students and 12 mentors across two workshops, we constructed a comprehensive dataset of 2,362 interaction sequences between students and mentors. These interactions were carefully coded into 23 specific student states, grouped under eight categories, and 20 mentor strategies, categorized under five guidance categories.

The dataset captures the dynamic nature of real-world classroom interactions, reflecting the diverse states of students and the adaptive strategies used by mentors. For instance, when a student shows confusion about a task's goals, a mentor might respond by restating the task and providing role guidance. Such examples demonstrate the variability and contextual nuances in student-mentor dialogues, providing valuable data for simulating effective CPS guidance.

To train a robust classification model for the Mentigo agent, we expanded our dataset by involving additional educational experts to annotate the data. This process ensured consistency and reliability, with each dialogue segment—be it a student query, response, or mentor guidance—being carefully reviewed and annotated for accuracy. Each of the 23 student states and 20 mentor strategies was validated multiple times across various scenarios to capture a broad spectrum of classroom interactions. This extensive annotation process enhances the dataset's diversity and adaptability, enabling the Mentigo system to effectively handle different classroom situations and provide tailored guidance in real-time.

In summary, the constructed dataset of 2,362 interaction sequences, with detailed annotations for student states and mentor strategies, forms the foundation for developing a data-driven agent. This dataset equips the agent to dynamically recognize and respond to different learning states and instructional needs in CPS environments.

\subsubsection{Technical Implementation Details}
Our system utilizes the Streamlit framework, enabling rapid development of interactive web applications directly from Python scripts. We chose Streamlit for its ability to integrate frontend and backend development, simplifying our workflow.

Data management is handled through Google Sheets, serving as a serverless backend for storing user information, conversation logs, and other critical data in real-time. This approach allows for flexibility and ease of scalability.

The conversational AI capabilities are powered by OpenAI's GPT-4o API (version 2024-08-06), complemented by Langchain for enhancing conversation structure and flow. This integration ensures high-quality, context-aware interactions.

Deployed on a cloud server, the system is accessible via web browsers, facilitating seamless user access as detailed Fig.~\ref{fig:UI}. Operational specifics and main prompts are shown in Fig. \ref{fig:Controller} and Fig.\ref{fig:mentor}.

\section{User Study}
We conducted a controlled user study under a within-subject design to evaluate the effectiveness and usability of Mentigo in supporting middle school students during CPS tasks. This study aimed to understand how Mentigo’s contextual intelligent guidance affects user engagement, depth of reflection, and integration of feedback in creative problem-solving processes. We hypothesize that the real-time personalized guidance provided by Mentigo will be perceived as a valuable tool, enhancing students' reflection and ideation during CPS tasks.

\subsection{Participants}
We recruited 12 middle school students aged 12 to 14 (6 males and 6 females) for this study (shown in Table 4). Participants had diverse backgrounds in AI tool usage, CPS experience, and computer skills. They were recruited through social media outreach, and all participants' parents provided informed consent. We employed a randomized within-subject experimental design, allowing each participant to interact with both a control group (using a baseline system) and an experimental group (using the Mentigo system). This design enabled a direct comparison of user experiences with and without Mentigo's intelligent guidance.

We implemented several ethical measures to ensure the safety and comfort of child participants. Informed consent was obtained from both the children and their parents, with clear explanations of the study’s purpose and procedures. Before each session, we reconfirmed the children’s willingness to participate and reminded them of their right to withdraw at any time. A researcher was present during all sessions to monitor well-being and provide support if needed. We also verified with parents that their children had no relevant mental health concerns and clearly communicated that the AI agent was not a real person. The study was approved by the university's Institutional Review Board (IRB).

\begin{table*}[ht]
\small
\centering
\caption{Participant's Information}
\begin{tabular}{ccccccc}
\hline
\textbf{ID} & \textbf{Gender} & \textbf{Age} & \textbf{Grade} & \textbf{CPS Experience} & \textbf{AI Experience} & \textbf{Computer Usage} \\
\hline
P1          & Boy             & 13           & 7              & 1                       & 1                      & 3                       \\
P2          & Boy             & 12           & 7              & 1                       & 2                      & 3                       \\
P3          & Boy             & 14           & 7              & 2                       & 1                      & 2                       \\
P4          & Girl            & 12           & 7              & 2                       & 1                      & 2                       \\
P5          & Girl            & 13           & 8              & 2                       & 2                      & 3                       \\
P6          & Girl            & 13           & 8              & 1                       & 1                      & 1                       \\
P7          & Girl            & 12           & 7              & 1                       & 1                      & 1                       \\
P8          & Boy             & 14           & 9              & 2                       & 2                      & 2                       \\
P9          & Boy             & 14           & 8              & 1                       & 2                      & 3                       \\
P10         & Girl            & 13           & 8              & 2                       & 2                      & 2                       \\
P11         & Boy             & 12           & 7              & 2                       & 2                      & 1                       \\
P12         & Girl            & 12           & 7              & 2                       & 2                      & 2                      
\\
\hline
\end{tabular}
\end{table*}

\subsection{Study Procedure and Protocol}

We adopted a within-subject experimental design for this user study. Participants were asked to complete CPS tasks under two conditions in a counterbalanced and randomized order: the system condition (using our fully designed educational agent with advanced features) and the baseline condition (using a basic version of the agent with simple prompts and a consistent interface). Two different CPS tasks were prepared for participants:

\begin{itemize}
    \item \textbf{Smart Home Task: }This task focuses on developing innovative solutions for smart home systems. Participants were asked to identify problems in current smart home technologies and propose creative solutions to improve user experience, safety, and sustainability.
\end{itemize}

\begin{itemize}
    \item \textbf{Low-Carbon Campus Task: }This task aimed to encourage participants to design initiatives or strategies to promote low-carbon practices in a school campus setting, considering factors such as energy efficiency, waste management, and engagement of students and staff.
\end{itemize}

For each task, participants followed a structured process that included gathering information, brainstorming solutions, and completing a learning report outlining their proposed solutions and the rationale behind them.

The study sessions were conducted remotely via Tencent Meeting. Participants were first introduced to the background and purpose of the study, completed a demographic survey, and read the consent form. After obtaining their consent, participants were guided through an introduction to the basic functionalities of the two agents and the study process. The study consisted of several phases:

\begin{enumerate}
    \item Introduction and Familiarization with the System (~10 minutes): Participants were introduced to the study's objectives and guided through the educational agent’s basic features, including interface navigation, task templates, and system guidance. For those familiar with similar tools, a quick refresher ensured all participants were comfortable using the system.

    \item Pre-Test Questionnaire (~5 minutes):
After familiarization, participants completed a pre-test questionnaire designed to measure their prior knowledge and skills related to CPS. This served as a baseline for evaluating changes in their CPS skills throughout the study.

    \item Task Sessions: System Condition and Baseline Condition (30-40 minutes each):
Participants were randomly assigned to complete the two tasks—Smart Home Task and Low-Carbon Campus Task—under two different conditions: the system condition using Mentigo and the baseline condition. The order of the conditions was counterbalanced to avoid learning effects.
 
  \textbf{System Condition Task (Mentigo):} In this condition, participants used Mentigo, the fully designed educational agent with advanced features such as dynamic prompts, reflective questions, and adaptive feedback. They had 30 minutes to complete the task with Mentigo's guidance, followed by completing a structured learning report.
 
  \textbf{Baseline Condition Task:} Participants used a simplified system version with basic prompts and a consistent interface, lacking the advanced features of Mentigo. After a brief introduction, they had 30 minutes to complete the second task, following the same procedure as in the system condition. 
  
    \item Post-Test Questionnaire for Each Condition (~5 minutes each):
After completing each task, participants filled out a post-test questionnaire to evaluate the system's usability, perceived cognitive load, and mastery of CPS skills. This feedback was crucial for comparing the effectiveness of Mentigo with the basic agent setup and understanding how the advanced features influenced user experience and learning outcomes.

    \item Final Interview (~20 minutes): After completing both conditions, participants took part in a final in-depth interview to provide qualitative insights on their experiences with both the Mentigo and baseline systems. The interview focused on their thought processes, challenges encountered, and preferences between the two systems. This feedback was vital for understanding user interactions and identifying key areas for enhancing the system’s design and functionality. 
\end{enumerate}

This combined procedure and protocol provided a comprehensive framework for evaluating the educational agent’s effectiveness and usability across different CPS contexts.

\subsection{Data Collection and Analysis}

Our data came from four primary sources: 1) pre-test and post-test questionnaires to assess knowledge gain, system usability, and task load, 2) learning reports created by participants during tasks to evaluate task performance outcomes, 3) screen recordings and system logs to capture student-system interaction data, and 4) in-depth interviews to gather qualitative feedback on user experiences. The pre- and post-test data were used for comparing learning outcomes and cognitive load between the Mentigo and baseline conditions, while the learning reports, interaction data, and interview data were used to analyze task performance, user engagement, and areas for system improvement.

\begin{table*}[ht]
\small
\centering
\caption{Manual Coding Scheme of System Data}
\begin{tabular}{p{3cm}ll}
\hline
Categories & Types & Description \\
\hline
\multirow{3}{3cm}{Speech Type} 
 & Positive & Proactively suggest solutions to problems; proactively ask questions.\\
 & Neutral & Express confusion or guide requests to AI\\
 & Negative & Request AI to provide direct answers; simply copy the AI's responses.\\
\hline
\multirow{6}{6cm}{Cognitive Level} 
 & Remembering& Reciting and memorizing labels\\
 & Understanding& Relating and organizing previous knowledge\\
 & Applying& Applying information into a new situation\\
 & Analyzing& Drawing connections among ideas\\
 & Evaluation& Examining information and make judgement\\
 & Creation& Creating new ideas using what has just been learned\\
\hline
\end{tabular}
\end{table*}

\subsubsection{Pre- and Post-Test Data}

To measure participants' knowledge gain, system usability, and cognitive load, pre-test and post-test questionnaires were administered before and after each task session. The pre-test assessed participants' baseline familiarity with CPS concepts, while the post-test evaluated system usability (specifically for the Mentigo condition), cognitive load, and CPS skills mastery. Knowledge gain was determined by comparing pre- and post-test results, reflecting changes in participants' self-rated knowledge on CPS tasks.

Descriptive statistics were used to summarize participants' demographics and baseline characteristics for data analysis. Paired-sample t-tests were conducted to compare cognitive load (measured using the NASA Task Load Index (TLX)~\cite{hart1988development}) and system usability between the Mentigo and baseline conditions. Repeated measures ANOVA were used to identify significant differences in knowledge gain and task load, providing insights into the educational agent's effectiveness and cognitive demands associated with each condition.

\subsubsection{Task Performance Data}

Participants' task performance was evaluated using learning reports created during each condition (Mentigo and baseline). These reports provided insights into participants' ability to apply CPS techniques, the quality of their solutions, and their reasoning processes.

To assess these reports, we developed an evaluation matrix for AI-involved CPS tasks based on Urban et al.~\cite{urban2024chatgpt}. The matrix included four dimensions: 1) Quality, 2) Elaboration, 3) Originality, and 4) Human-like Expression (see supplementary materials for details). Three researchers independently rated the reports according to these criteria, ensuring reliability through inter-rater agreement.

The scores were analyzed using descriptive statistics to summarize performance in each dimension and non-parametric tests, such as the Mann-Whitney test, to compare differences between the Mentigo and baseline conditions. This provided insights into how effectively the educational agent supported CPS skills development.

\subsubsection{Student-System Interaction Data}

Screen recordings and system logs were collected to capture participants' interactions with the educational agent during the study sessions. These data sources included metrics such as time spent on tasks, prompt usage frequency, and action sequences. This information helped analyze user engagement, behavior patterns, and how participants navigated the system and utilized feedback to improve their CPS processes.

For analysis, sequence analysis and descriptive statistics were applied to the screen recordings and system logs to understand user behavior and engagement patterns. Metrics such as time spent on tasks, frequency of prompt usage, and interaction paths were examined to assess how participants used the agent’s features (see Table 5). The analysis aimed to identify common strategies, variations in system use, and how these factors contributed to task outcomes.

\subsubsection{Interview Data}

In-depth interviews were conducted after participants completed both conditions to gather qualitative feedback on their experiences with the educational agent. The interviews explored participants' thought processes, the impact of system guidance, the challenges faced, and their overall preferences between the two conditions. The interviews were recorded, transcribed, and analyzed to identify themes and areas for potential system improvement.

Thematic analysis was employed to analyze the qualitative data from the interviews. The coding process was iterative, focusing on refining themes that captured participants' experiences, perceived value, and suggestions for improvement. The results provided a comprehensive understanding of user experiences and were used to inform future iterations of the educational agent system (see supplementary materials for the Interview Protocol Outline).

By combining these diverse measurement and analysis methods, our study provided a comprehensive assessment of Mentigo's effectiveness and usability in mentoring middle school students' CPS tasks. This mixed-methods approach enabled a thorough understanding of both the measurable outcomes and the subjective experiences of participants.

\subsection{Expert Review and Feedback}
To further validate the findings and gather insights into the system’s effectiveness, we conducted an expert review and feedback session. We recruited five experts, including experienced secondary school teachers, a specialist in educational research, and a mentor with extensive experience in guiding middle school students. These experts reviewed the data generated from the study, such as students' learning reports, task outcomes, and dialogue records with the two systems.

The experts evaluated the system's educational performance, usability, and impact on students’ CPS skills. Following their review, they participated in a semi-structured interview to provide detailed feedback on how the system supported student engagement and learning outcomes. They highlighted the system’s strengths and identified areas for improvement, offering practical suggestions based on their experience with middle school students.

The insights from these expert sessions were crucial for refining our analysis of the two system performances. The feedback also provided actionable guidance for future development to enhance the system’s support for CPS and improve student learning experiences in real-world educational settings.

\section{Findings}

\subsection{Evaluation of System Usability and Effectiveness}

The Mentigo system is a data-driven intelligent agent, which is built on a robust foundation of over 80 hours of dialogue data collected from real CPS educational settings. It guides students through a CPS task using a six-stage process. At each stage, the system sets clear task requirements and progression standards, using corresponding guidance strategies and generative AI to ensure a personalized learning experience that adapts to each student's state and stages. Generally, students spent more time on tasks (M = 43.7 mins vs. 24.7 mins) and had more conversation rounds (M = 24.0 vs. 13.2) with Mentigo compared to the Baseline system, reflecting a more thorough and reflective learning process. 

The Mentigo system demonstrated a high level of usability, with a System Usability Scale (SUS) score of 88.54, comparable to the Baseline system's score of 89.79. While both systems are highly usable, further analysis reveals significant differences in key performance dimensions that highlight Mentigo's strengths in guiding students through CPS tasks (shown in Table 6). The system excels in providing clarity of goals (\textit{p} = 0.051) and delivering comprehensive guided strategies (\textit{p} = 0.032), which enhances student engagement and promotes deeper learning. For instance, P5 said, \textit{''The system A (Mentigo) actively guides you to complete the tasks, providing a structured and purposeful approach that helps us focus on our goals.''} Students can follow its model to explore and learn specific knowledge, although it still requires students to do information integration and research, which encourages deeper engagement and learning. Mentigo outperformed the Baseline system in knowledge and skills acquisition (\textit{p} = 0.019), effectively helping students learn to articulate issues and gain CPS knowledge. This aligns with the qualitative feedback where students expressed high appreciation for Mentigo's clarity in outlining tasks at each stage and guiding them. P1 mentioned, \textit{"System A (Mentigo) is more like a project mentor. It provides more guidance and leads you step-by-step, almost like a teacher guiding you through the process." } 

\begin{table}[ht]
\small
\centering
\caption{Comparison of Condition System and Baseline System Across Different Dimensions}
\begin{tabular}{lllll}
\hline
Dimension & Mentigo & Baseline & Value & $p$ \\
\hline
Helpfulness & M=8.83 & M=8.75 & t=0.17 & 0.867\\
Goal Clarity & M=9.17 & M=7.42 & t=2.07 & 0.051$\wedge$\\
Guided Strategy & M=9.17 & M=8.17 & t=2.28 & 0.032*\\
Personalization & M=9.17 & M=8.83 & t=0.98 & 0.339\\
Knowledge and Skills Acquisition & Mdn=9 & Mdn=7.5 & U=112 & 0.019*\\
Emotional Support & M=8.33 & M=7.42 & t=1.18 & 0.25\\
\hline
\end{tabular}
\end{table}

Mentigo's core strength lies in its ability to dynamically identify and respond to over 23 distinct student states, such as "unclear task," "overly divergent thinking," and "surface-level thinking" (see Table 3 for details). The analysis of student state frequencies shows that states like ID 1 ("Silent"), ID 2 ("Goal Unclear"), and ID 23 ("Normal Progress/Active Questioning") are most frequently addressed, indicating these are common areas where students need support. Moderately frequent states, such as ID 10 ("Shallow Thinking"), ID 7 ("Limited Thinking"), and ID 13 ("Inadequate Information Integration Skills"), suggest that challenges in critical and creative thinking are also significant but less prevalent. Less common states, including ID 4 ("Goal Prioritization Confusion"), ID 17 ("Decision Uncertainty"), ID 18 ("Insufficient Intrinsic Motivation"), and ID 19 ("Lack of Confidence"), highlight Mentigo's capacity to provide targeted support for specific, less frequent student needs. This varied distribution of student states reflects Mentigo's adaptability and effectiveness in delivering personalized educational experiences tailored to diverse learning needs.

Similarly, the distribution of guided strategy IDs (see Table 2 for details) shows that strategies like ID 1 ("Task Follow-up and Allocation"), ID 15 ("Resource Guidance"), and ID 18 ("Encouragement and Motivation") are frequently employed, highlighting Mentigo's focus on these methods to effectively support students. Other strategies, such as ID 8 ("Probing Guidance"), ID 9 ("System Thinking Training"), and ID 13 ("Resource Guidance"), are also used often, indicating their targeted application for specific needs. Less frequent strategies, like ID 5 ("Creative Filtering and Focusing"), ID 11 ("Deep Thinking and Reflection"), and ID 19 ("Emotional Counseling"), demonstrate the system’s adaptability in addressing unique challenges. This varied use of instructional strategies underscores Mentigo's ability to provide personalized and flexible educational experiences tailored to diverse learning contexts. Overall, 

\begin{figure*}[h]
  \includegraphics[width=\textwidth]{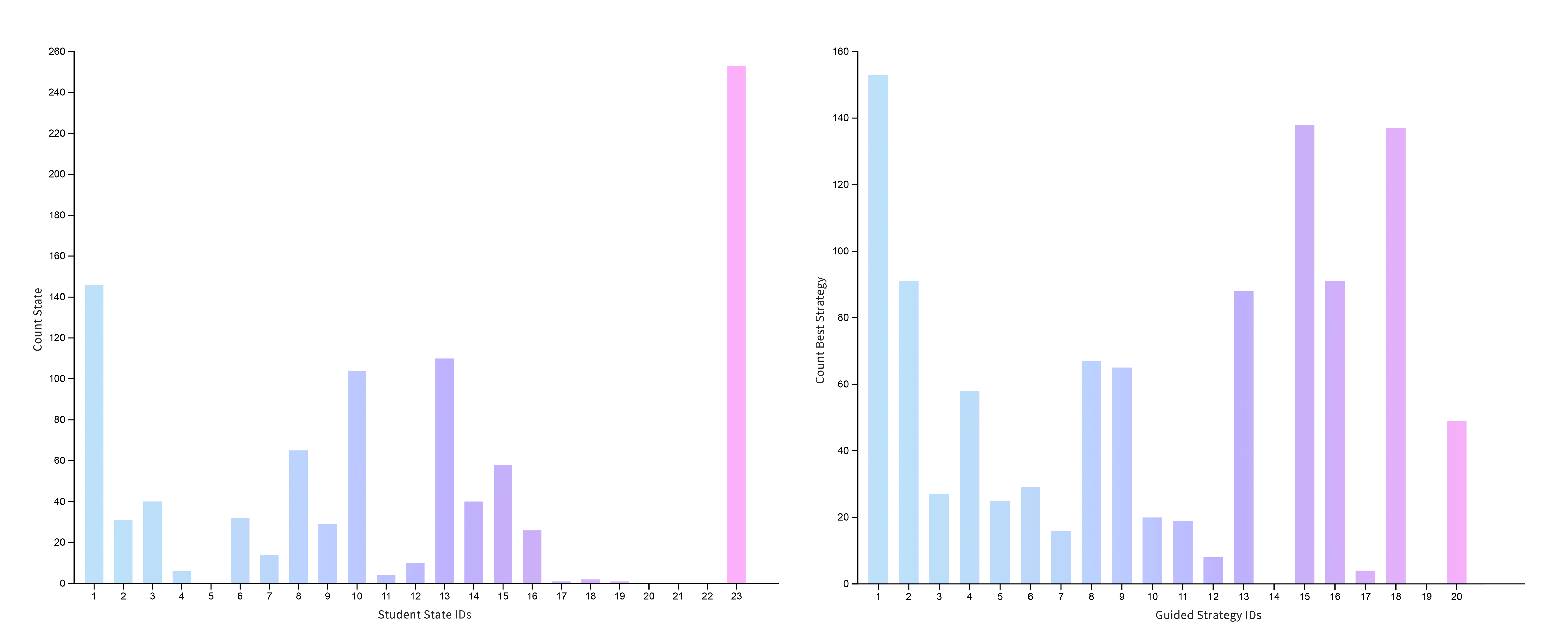}
  \caption{Distribution of Student States and Guided Strategies Frequencies}
  \label{fig:ID}
\end{figure*}

\subsection{Student CPS Knowledge Improvement and Performance}

Based on the analysis of learning outcomes, a Mann-Whitney U test was conducted to compare the pre-test results between the Mentigo and Baseline groups (as shown in Table 7). The results suggested that there were no significant differences between the two groups in their pre-test scores (\textit{p}= 0.51). This indicates that both groups started with similar levels of knowledge before the intervention.

For the Mentigo group, there was a notable improvement from the pre-test to the post-test (\textit{p}=0.006). The post-test scores (M = 6.67, SD = 2.19) were higher than the pre-test scores (M = 4.33, SD = 2.23), with a large effect size (Cohen's d = 0.944), suggesting substantial learning gains. In contrast, the Baseline group showed similar median scores between the pre-test (M = 4.92, SD = 2.61) and post-test (M = 5, SD = 1.91), though the effect size (Cohen's d = 0.036) indicates a medium difference, implying less change. When comparing the post-test scores between the Mentigo and Baseline groups,  the effect size (Cohen's d = 0.765) suggests a large difference favoring the Mentigo group, and the Mann-Whitney U test results indicated that this difference was statistically significant ( \textit{p}= 0.019). P3 stated, \textit{``I prefer the first system, Mentigo, because the baseline system often just tells you what to do directly, while Mentigo tells you the steps of thinking and guides you to think step-by-step.''} This feedback highlights how Mentigo facilitates deeper cognitive engagement by guiding users through a structured thinking process, which aligns with the observed improvement in knowledge levels. On the other hand, P3, P8, and P9 mentioned that \textit{"while the Baseline System is quicker to use and feels more like a tool that provides reference answers, it doesn’t give a strong sense of accomplishment."} They described the experience as using a tool that gives quick solutions rather than fostering deep understanding. In contrast, \textit{System A (Mentigo) helps us learn more methods and strategies and offers more guidance for critical thinking, leading to a richer learning experience and greater gains in knowledge and skills.} P12 added.

\begin{table}[ht]
\small
\centering
\caption{Pre-/post-test Results (mean and standardized deviation) of Mentigo and Baseline}
\begin{tabular}{lllll}
\hline
& Mentigo&  Baseline&  Effect size & $p$ \\
\hline
Pre-test& 4.33(2.23)& 4.92(2.61)& 0.19 & 0.51 \\
Post-test& 6.67(2.19)& 5(1.91)& 0.77* & 0.019*\\
Effect size&0.944**& 0.036\\
p-value& 0.006** & 0.9\\
\hline 
\end{tabular} 
\end{table}

Besides, we also evaluated the students' learning reports completed using the two systems across four dimensions. An independent samples t-test revealed a significant difference between the scores for Mentigo (\textit{M} = 13.43, \textit{SD} = 3.17) and the baseline system (\textit{M} = 10.74, \textit{SD} = 2.72), with \textit{p} = 0.016. This result suggests that students' performance, as reflected in their learning reports, was significantly higher when using Mentigo compared to the baseline system. This is consistent with what P12 and P10 observed: \begin{quote}" By using Mentigo, I learned more methods and strategies, and I gained more from it. It pushes me step-by-step to complete all the tasks, and the ideas provided are based on the questions I raised, which often gives me a moment of clarity and leads me to think more comprehensively."\end{quote} These comments suggest that Mentigo's tailored guidance and reflective approach lead to deeper understanding and more significant learning gains compared to the baseline system. P11 further added, \textit{"Mentigo gives many prompts that allow me to think for myself, which makes me more creative because it focuses more on my own answers and guides me to write step-by-step."} This insight indicates that Mentigo encourages more creative and independent problem-solving, enhancing the overall learning experience.

Overall, these results indicate that while both systems significantly impacted students' knowledge and performance, Mentigo provided a more structured and comprehensive learning experience that led to greater improvement in CPS knowledge and skills. The combination of quantitative data and qualitative insights suggests that Mentigo's approach to guiding students and encouraging deep thinking was more effective in enhancing their overall learning outcomes.

\subsection{Student Engagement Across Mentigo and Baseline Systems}
To analyze student engagement across the Mentigo and Baseline systems, we compared data on student-system interactions, including the number of dialogue turns and usage time. Additionally, we encoded the content of student-system dialogues to assess different aspects of engagement. Due to time constraints, two students did not complete the tasks in the last two stages of the experiment using our system; these incomplete data were excluded from the final analysis. While we measured task load across various dimensions and found no significant differences between the Mentigo and Baseline systems, our system did not increase cognitive load for students, and overall student engagement remained high. The comparison between the Mentigo and Baseline systems across different dimensions shows significant differences in several areas of student engagement: behavioral, cognitive, and emotional.

\paragraph{Behavioral Engagement}

At the behavioral level, we assessed students' task duration, the number of conversation rounds, and the word count of student speech while using both systems (as shown in Table 8 ). The results showed that students had significantly longer task durations (\textit{p} = 0.005) and more conversation rounds (\textit{p} = 0.015) when using Mentigo. The word count of student speech was also significantly higher in the Mentigo group compared to the baseline group (\textit{p} = 0.014). This observation aligns with our experimental findings, where students tended to think more deeply about the CPS questions while using Mentigo. The system's guided prompts encouraged them to complete each task step-by-step and to engage more proactively. This suggests that the structured guidance in Mentigo fosters higher levels of engagement and encourages students to think and participate more actively.

\begin{table*}[ht]
\small
\centering
\caption{Comparison of Mentigo and Baseline System Across Different Dimensions}
\begin{tabular}{p{3cm}lllll}
\hline
Categories & Variable & Mentigo & Baseline & Value & $p$ \\
\hline
\multirow{3}{3cm}{Conversation Metrics} 
&Task Duration (min) & M = 43.7 & M = 24.3& t = 3.19 & 0.005** \\
&Conversation Rounds & Mdn = 24.5& Mdn = 12.5& U=82.5& 0.015*\\
&Word Count & Mdn = 1438.5& Mdn = 290& U=83& 0.014*\\
\hline
\multirow{3}{3cm}{Speech Type} 
 & Positive & M = 15.9& M = 7.2& t = 3.29 & 0.0041** \\
 & Neutral & Mdn = 5.5& Mdn = 3.5& U=65.5& 0.254\\
 & Negative & M = 2.5& M = 2.7& t = -0.15 & 0.8812 \\
\hline
\multirow{6}{6cm}{Cognitive Level} 
 & Memory & M = 1.9& M = 1.5& t = 0.58 & 0.5717 \\
 & Understanding & M=2.4& M=2.1& t=0.3& 0.7692\\
 & Application & M = 5.1& M = 4.8& t = 0.13& 0.9017\\
 & Analysis & M = 4.3& M = 1.7& t = 2.16& 0.0444*\\
 & Evaluation & M = 3& M = 0.9& t = 2.64& 0.0166*\\
 & Creation & Mdn = 7.5& Mdn = 2& U=81.5& 0.0180*\\
\hline
\end{tabular}
\end{table*}

\paragraph{Cognitive Engagement}
For cognitive engagement, we coded the content of students' speech according to type (Positive, Neutral, Negative) and cognitive level, based on Bloom's Taxonomy (six levels). Significant differences were found in the higher-order cognitive levels—specifically, "Analysis," "Evaluation," and "Creation"—with Mentigo outperforming the baseline system (\textit{p} = 0.044, \textit{p} = 0.0166, and \textit{p} = 0.0180, respectively). This highlights how Mentigo, acting more like a mentor, encourages deeper cognitive engagement and supports the development of CPS skills through guided exploration and reflection. This highlights how Mentigo, acting more like a mentor, encourages deeper cognitive engagement and supports the development of CPS skills through guided exploration and reflection. P11 mentioned, \begin{quote}
    "When using the baseline system, it’s easy to lose track and not know what to do next. But Mentigo has a more structured flow, making it less likely to go off-topic, and the guidance provided is very comprehensive, which leads to more complete thinking."
\end{quote} This statement underscores Mentigo's ability to provide a structured and comprehensive framework that helps students maintain focus and think more broadly, contrasting with the more direct answer-providing nature of the baseline system.

\paragraph{Emotional Engagement}
In terms of emotional engagement, although quantitative results showed no statistically significant differences in emotional support between the two systems, the interview data provided further insights. Many students felt that Mentigo was more like conversing with a human, using simpler language, emojis, and interactive encouragements. For instance, P7, P8, and P9 noted that \textit{"Mentigo feels more like a peer, providing encouraging messages and engaging in more interactive dialogues. In contrast, the baseline system feels more like a reserved friend who responds when asked but doesn't initiate conversations."} This suggests that Mentigo provides a more supportive and engaging emotional environment, particularly suitable for middle school students, whereas the baseline system seems more like a hyper-rational expert that lacks warmth and adaptiveness in communication.

These findings indicate that Mentigo provides a more comprehensive and effective environment for student engagement across behavioral, cognitive, and emotional dimensions. It promotes active participation, higher-order thinking, and emotional connection, making it more suitable for fostering CPS skills and creativity among middle school students. The baseline system, while functional, appears to focus more on information retrieval and direct answers, with less emphasis on interactive and personalized learning experiences.

\subsection{Expert's Review and Feedback}

We conducted interviews with five experts to gather insights into students' cognitive changes, AI's guidance strategies, and interactions with different systems (Mentigo vs. Baseline). The experts reviewed dialogue data between students and the systems, along with learning report data, and provided comprehensive feedback from an educational perspective. Experts noted significant differences in how students interacted with the Mentigo system compared to the Baseline system. E1 observed, \textit{"It often seemed that students treated the baseline system more like a tool to command and directly get answers rather than engaging in deep thinking or dialogue."}This suggests that the structured guidance in Mentigo fosters higher levels of engagement and encourages students to think and participate more actively. When reviewing the dialogues of P1 using Mentigo and baseline system, E3 highlighted that as the conversation progresses, the students show "more comprehensive expressions and transfer previously mentioned content into new contexts. In contrast, the baseline system felt more like an information retrieval tool with limited cognitive advancement. E1 added, \textit{"The baseline system is more like a didactic expert who provides answers but does not engage with students at a deeper level."} This difference highlights how Mentigo, acting more like a mentor, encourages deeper cognitive engagement and supports the development of CPS skills through guided exploration and reflection.

This observation was echoed by E2, who elaborated, \textit{"Mentigo offers more patient encouragement, like a cheerful artist friend who is more expressive and proactive, even using emojis to ask questions."} This supports the feedback from students who noted that Mentigo provided comprehensive guidance and multiple perspectives for thinking. E5 mentioned, \textit{"The guidance was simple yet offered various perspectives and space for thinking,"} suggesting that the system's ability to guide through structured steps supports its quantitative superiority in fostering cognitive engagement. In contrast, E4 commented that the Baseline System's approach often felt like a "question-and-answer session," lacking the same depth of guidance. Experts praised Mentigo's ability to provide structured guidance and maintain a comprehensive framework that helps students remain focused and think more deeply. 

\begin{figure*}[!htp]
    \centering
    \includegraphics[width=1.0\linewidth]{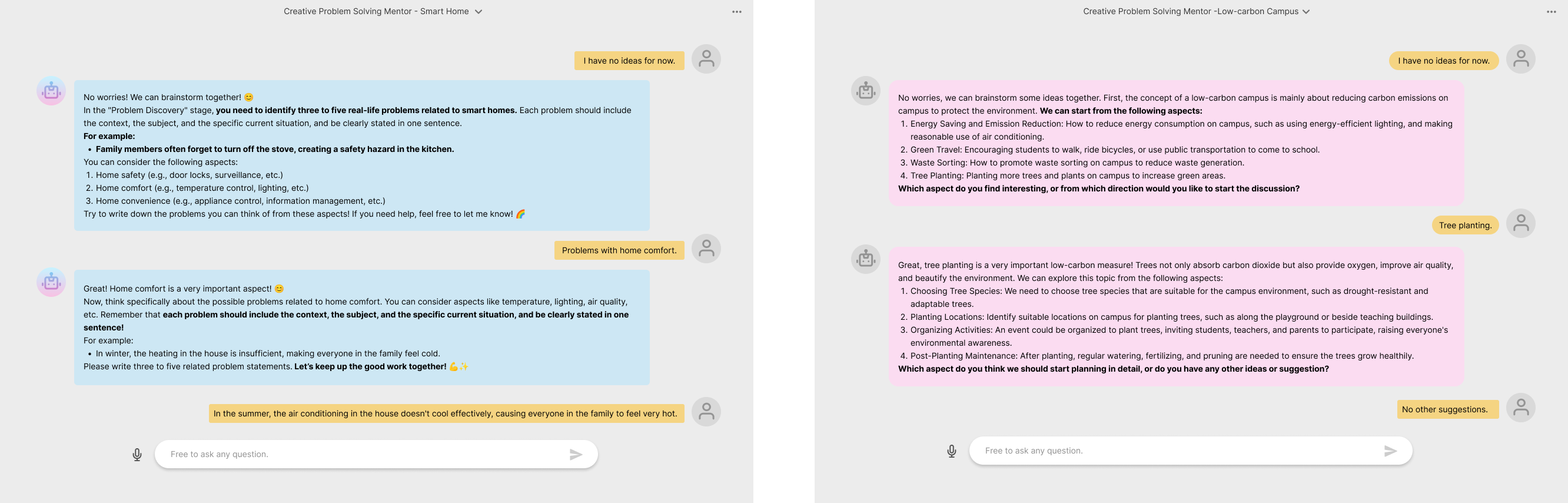}
    \caption{Illustration of Student P1's Dialogues with Mentigo (Left) and the Baseline System (Right) at the Same Stage of Different Tasks}
    \label{fig:UI}
\end{figure*}

E5's feedback supports this observation: When P1 interacted with both systems at the beginning of the tasks, the differences in their guidance became evident. System A (Mentigo) provides clear direction at the start of each stage, explaining the task objectives and guiding students to discover more problems from multiple perspectives based on their input. It also prompts students to clarify their problem statements as required. Meanwhile, when P1 interacted with System B, it generated multiple answers based on the provided keywords for the student to choose from. This often led to students quickly adopting the AI-generated answers and ending the conversation prematurely. An excerpt of these interactions is shown in Fig.~\ref{fig:UI}. This highlights Mentigo's role in scaffolding the learning process and helping students build their CPS skills step-by-step.

Experts also highlighted the contrast between the two systems' approaches: while Mentigo was described as a "mentor," the Baseline system was perceived as a "knowledge authority" that simply provides information. E3 pointed out that \textit{"System B (Baseline) felt like an over-rational tool; it lacked the 'human touch,' making it difficult for students to relate and engage deeply."} This difference impacted students' engagement levels, with Mentigo being more effective in fostering both cognitive and emotional engagement. Besides, as for the emotional support, E1, E3, and E4 stated that Mentigo was more like conversing with a human, using simpler language, emojis, and interactive encouragements. For instance, E4 noted that \textit{"Mentigo feels more like a peer, providing encouraging messages and engaging in more interactive dialogues."} In contrast, the Baseline system feels more like a reserved friend who responds when asked but doesn't initiate conversations. This suggests that Mentigo provides a more supportive and engaging emotional environment, particularly suitable for middle school students, whereas the Baseline system seems more like a hyper-rational expert that lacks warmth and adaptiveness in communication. Besides, expert feedback suggests a significant opportunity to develop a teacher interface for real-world applications. E2, E3, E5 both highlight Mentigo offers valuable insights into students' thinking processes and progress, which could be further enhanced by making these processes visible to educators in real time. E2 stated, \begin{quote}I think a teacher interface could allow educators to monitor students' states, visualize their thought processes, and provide formative assessments based on real-time data. Such a feature would enable teachers to intervene more effectively, offer targeted support, and refine AI-driven guidance strategies, creating a more integrated and dynamic educational environment.\end{quote}

Overall, the experts' feedback suggests that the Mentigo system, with its structured and personalized guidance, offers a more effective and engaging environment for students compared to the Baseline system. It promotes cognitive development, encourages deeper engagement, and creates a supportive emotional atmosphere that is crucial for fostering creativity and problem-solving skills. The insights gathered from the experts provide valuable directions for further refining the AI's guidance strategies to better support diverse student needs and learning styles.

\section{Discussion} 

\subsection{Enhancing Engagement and Learning Outcomes in Creative Problem Solving with Mentigo}

Creative Problem Solving (CPS) is a critical skill in education, fostering students’ ability to think creatively, engage in reflective dialogue, and develop innovative solutions to complex real-world challenges. CPS is known to significantly enhance students' cognitive skills, creativity, and problem-solving abilities, making it an essential component of modern education~\cite{samson2015fostering, treffinger1995creative}. However, guiding students through CPS can be resource-intensive, requiring continuous support and feedback that traditional classroom settings often struggle to provide.

Mentigo addresses these challenges by acting as an AI mentor that fosters significantly higher levels of student engagement— behavioral, emotional, and cognitive—compared to traditional systems. Key findings from our study indicate that students using Mentigo were more actively engaged in their learning, demonstrating longer task durations, more conversation rounds, and richer vocabulary in their responses. This suggests that Mentigo’s structured guidance and adaptive feedback help students maintain focus and engage more deeply with their tasks, promoting higher levels of cognitive engagement and reflective thinking~\cite{samson2015fostering, lavender2006creative, yu2017mentoring}. Some students took the initiative to suggest additional steps or ask new questions, such as, "What if we consider another scenario?" or "Could we explore an alternative solution?" These interactions indicate students’ growing willingness to engage in higher-order thinking and explore multiple perspectives, key components of active learning and cognitive development~\cite{neumann2020intelligent, ashoori2007mentor}.

Overall, the use of Mentigo transformed the traditional learning environment into a more interactive and engaging space for students. Its mentoring approach, characterized by structured guidance, personalized feedback, and an emphasis on reflective thinking, successfully promoted active participation and deeper cognitive engagement among students. By supporting individual exploration and encouraging students to think critically and creatively, Mentigo has demonstrated its potential as an effective educational tool to mentor students and enhance their overall learning outcomes.

\subsection{Balancing Structured Guidance and Personalized Feedback for Enhanced CPS Skills}
The findings indicate that Mentigo’s structured task management system, combined with its diverse range of guided strategies, effectively enhances students' clarity, compliance, and development of CPS skills. By breaking down the CPS process into clear, manageable stages and providing explicit instructions, Mentigo helps students focus on task objectives and progress systematically, reducing cognitive overload and supporting better learning outcomes~\cite{samson2015fostering, eberle2021cps}. However, while structured management improves task focus, it can limit creative exploration if not balanced with open-ended opportunities. To address this, Mentigo integrates varied guidance strategies that encourage exploration and creativity, aligning with research on fostering critical thinking through diverse prompts~\cite{neumann2020intelligent, gizzi2022creative}. Expert feedback suggests refining the system to adaptively switch between guidance strategies based on real-time student feedback, ensuring a dynamic balance between structure and flexibility. Future designs could incorporate more adaptive task management mechanisms that adjust dynamically to a learner’s progress and needs.

Additionally, the study highlights the importance of personalized feedback that responds to students' dynamic states. Mentigo’s monitoring of cognitive and emotional states to deliver tailored feedback has proven effective in maintaining engagement and promoting deeper learning. Students reported feeling more supported and understood, motivating them to engage more deeply with the tasks. This aligns with prior research on the value of personalized, context-aware feedback in learning environments~\cite{lavender2006creative, ashoori2007mentor}. Future improvements could explore integrating more multimodal feedback options to accommodate different learning preferences and further enhance engagement.

\subsection{Encouraging Deeper Cognitive and Emotional Engagement Through Adaptive, Empathetic Interactions}

Mentigo shows a strong capability to foster deeper cognitive engagement, demonstrated by the increased instances of higher-order thinking, such as analysis, evaluation and creation. Its adaptive prompts and scaffolding strategies encourage reflection, critical thinking, and the exploration of alternative solutions, supporting the development of creative problem-solving skills. This aligns with educational theories advocating for adaptive learning environments to stimulate critical thinking and creativity~\cite{treffinger1995creative, hudson2013mentoring}. However, some students struggled with the level of challenge presented by the system. Future development should consider more nuanced adaptive algorithms that can fine-tune difficulty levels based on each student’s progress and comfort, ensuring a more personalized learning experience.

Empathetic interactions are also a central design focus of Mentigo, aimed at fostering emotional engagement and creating a supportive learning environment for middle school students. While quantitative results did not show a significant difference in emotional support between the groups, expert interviews suggested that Mentigo provided a "warmer" and more supportive experience compared to the baseline system. This perception aligns with research on the benefits of empathetic AI in enhancing emotional well-being and motivation~\cite{yu2017mentoring, gizzi2022creative}. Experts highlighted the need for more sophisticated emotion recognition to make responses more context-aware and personalized. Future development should focus on refining the AI’s language and interactions to be more relatable and supportive, leveraging real-time emotion detection to better meet the emotional needs of students and foster a positive learning experience.

\subsection{Limitation and Future Work}
While Mentigo shows potential in enhancing student engagement in CPS tasks, there are several limitations to consider. The system's stability issues, such as occasional lags and response delays, could hinder user experience and reduce engagement. Additionally, the study's limited scale, involving a specific group of middle school students in controlled conditions, may affect the generalizability of the results to broader educational contexts. Furthermore, while the system provides adaptive guidance, its current ability to dynamically adjust the difficulty level and feedback based on real-time student interactions could be refined to better cater to diverse learning needs.

For future work, it is essential to deploy Mentigo in real-world classroom settings to evaluate its effectiveness amidst the complexities of authentic educational environments. This would provide deeper insights into how the system supports diverse student populations and adapts to varying classroom dynamics. Another promising direction is to enhance the system's capabilities for collaborative learning, allowing for group problem-solving tasks where AI mentors can facilitate peer interactions. Additionally, future research could explore integrating AI mentors more closely with human teachers, creating seamless teacher-AI-student collaboration. Developing intuitive teacher interfaces for monitoring student progress and providing real-time interventions could further enhance educational outcomes and foster a more cohesive and dynamic learning environment.

\section{Conclusion}
We presented Mentigo, an AI-driven educational agent designed to enhance student engagement through structured guidance, personalized feedback, and empathetic interactions. By leveraging large language models, Mentigo provides an adaptive learning experience tailored to CPS tasks. To evaluate its effectiveness, we conducted an with-in experiment study comparing Mentigo with a baseline system, revealing that students using Mentigo were more engaged behaviorally, cognitively, and emotionally. These findings suggest that AI-driven mentoring systems like Mentigo can significantly enrich the learning experience in educational settings. We hope this work contributes to the HCI and educational technology field by demonstrating the potential of AI systems to provide personalized and adaptive mentor that fosters deeper learning and engagement.


\bibliographystyle{ACM-Reference-Format}
\bibliography{main}


\end{document}